\documentclass[a4paper,twoside,usernatbib]{article}
\newcommand{\affil}[1]{$^{\rm #1}$}
\usepackage[authoryear]{natbib}
\bibpunct{(}{)}{;}{a}{}{,}
\usepackage{graphicx}
\date{} 
\title{\large\bf\flushleft The PULSE@Parkes project: A new observing technique for long-term pulsar monitoring}
\author{\parbox{\textwidth}{\flushleft
\vspace{-0.5cm}
{\it G. Hobbs\affil{A}, R. Hollow\affil{A}, D. Champion\affil{A}, J. Khoo\affil{A}, D. Yardley\affil{A,B}, M. Carr\affil{A}, M. Keith\affil{A}, F. Jenet\affil{C}, S. Amy\affil{A}, M. Burgay\affil{D}, S. Burke-Spolaor\affil{A,E}, J. Chapman\affil{A}, L. Danaia\affil{F}, B. Homewood\affil{G}, A. Kovacevic\affil{H}, M. Mao\affil{A,I,J}, D. McKinnon\affil{F}, M. Mulcahy\affil{A}, S. Oslowski\affil{A,E} and W. van Straten\affil{E}}\\
\vspace{0.4cm}
{\small \affil{A}\,Australia Telescope National Facility, CSIRO, P.O. Box 76, Epping, NSW 1710 Australia.}\\
{\small \affil{B}\,Sydney Institute for Astronomy (SIfA), School of Physics, The University of Sydney, NSW 2006, Australia}\\
{\small \affil{C}\,Center for Gravitational Wave Astronomy, University of Texas at
Brownsville, 80 Fort Brown, Brownsville, TX 78520}\\
{\small \affil{D}\,Universit\`a di Cagliari, Dipartimento di Fisica, SP Monserrato-Sestu km 0.7, 09042 Monserrato (CA), Italy}\\
{\small \affil{E}\,Centre for Astrophysics and Supercomputing, Swinburne University of Technology, P.O. Box 218, Hawthorn VIC 3122, Australia} \\
{\small \affil{F}\,Charles Sturt University, Bathurst, Australia} \\
{\small \affil{G}\,Braemar College, Woodend, Victoria, Australia} \\
{\small \affil{H}\,Department of Physics, Macquarie University, Sydney, NSW 2109, Australia}\\ 
{\small \affil{I}\,Anglo-Australian Observatory, PO Box 296, Epping, NSW, 1710, Australia}\\ 
{\small \affil{J}\,School of Mathematics and Physics, University of Tasmania, Private Bag 37, Hobart, 7001, Australia}}}
\begin{document}
\twocolumn[
\begin{changemargin}{.8cm}{.5cm}
\begin{minipage}{.9\textwidth}
\vspace{-1cm}
\maketitle
%
%
\small{\bf Abstract:}
The PULSE@Parkes project has been designed to monitor the rotation of radio pulsars over time spans of days to years. The observations are obtained using the Parkes 64-m and 12-m radio telescopes by Australian and international high school students.  These students learn the basis of radio astronomy and undertake small projects with their observations. The data are fully calibrated and obtained with the state-of-the-art pulsar hardware available at Parkes.  The final data sets are archived and are currently being used to carry out studies of 1) pulsar glitches, 2) timing noise, 3) pulse profile stability over long time scales and 4) the extreme nulling phenomenon.  The data are also included in other projects such as gamma-ray observatory support  and for the Parkes Pulsar Timing Array project.  In this paper we describe the current status of the project and present the first scientific results from the Parkes 12-m radio telescope. We emphasise that this project offers a straightforward means to enthuse high school students and the general public about radio astronomy while obtaining scientifically valuable data sets.  

\medskip{\bf Keywords:} pulsars: general

\medskip
\medskip
\end{minipage}
\end{changemargin}
]
\small

\section{Introduction}

Long-term and frequent observations of radio pulsars have led to a large number of important and diverse astrophysical results.  For instance, long-term timing experiments have generated studies of basic physics such as providing stringent tests of the theory of general relativity (Kramer et al. 2006\nocite{ksm+06}), and have made important astrophysical discoveries including the first extra-Solar planet (Wolszczan \& Frail 1992\nocite{wf92}). Pulsar timing experiments have also led to an improved understanding of pulsars and their surroundings including 1) an understanding of the pulsar velocity distribution (Hobbs et al. 2005\nocite{hllk05}), 2) the discovery of pulsar glitch events (e.g. Shemar \& Lyne 1996\nocite{sl96}) and spin-down irregularities (e.g. Hobbs, Lyne \& Kramer 2006\nocite{hlk06}) and 3) the first evidence of ion-neutral damping in the interstellar medium (You et al. 2007\nocite{yhc+07}).  Existing long-term pulsar timing programs are attempting even more ambitious experiments, such as making the first direct detection of gravitational wave signals (e.g. Manchester 2006\nocite{man06}).  Pulsar timing observations are also essential for gamma-ray telescopes such as the Gamma-ray Large Area Space Telescope (previously known as GLAST and now re-named Fermi) and the Astro-rivelatore Gamma a Immagini LEggero (AGILE).  In order for these observatories to detect and study the emission from pulsars they need accurate pulsar timing models that can only be supplied from radio timing \citep{sgc+08}.

The basic process of pulsar timing has been described numerous times in the literature.  In brief, pulsar observations give us measurements of pulse times-of-arrival (TOAs) at an observatory. Computer software (for instance, \textsc{tempo2}; Hobbs, Edwards, Manchester 2006\nocite{hem06}, Edwards, Hobbs \& Manchester, 2006\nocite{ehm06}) is required for carrying out the complex calculations to convert the measured TOAs to the proper time of emission. The calculations include converting the measured site arrival times to Solar System barycentric arrival times and adding excess propagation delays caused by the interstellar medium.  The \textsc{tempo2} software is used to compare the derived time of emission with a pulsar model to form ``timing residuals'',  the deviations between the observed TOAs and the model predictions.  For a perfect pulsar model and random receiver noise, these timing residuals will be uncorrelated and will have a flat spectrum.  The pulsar model can be improved by model-fitting to the timing residuals.  However, any remaining residuals correspond to unmodelled physics affecting the measured TOAs.  These may indicate calibration or instrumental errors or may be caused by unmodelled binary companions to the pulsar (such as planetary companions), or the effect of gravitational wave signals on the pulsar and/or the Earth.

Long-term pulsar timing experiments rely on the dedication of the observing team and the willingness of the time allocation committee to allocate telescope time over many years often without immediate scientific results.  This is particularly difficult for telescopes such as the 64-m diameter Parkes radio telescope which are not only oversubscribed, but also require that at least two observers be present at the telescope for every observation.

The importance of public outreach work in present and future large-scale projects is clear. Even though some major radio astronomy projects have specific funding for outreach it often takes the form of websites describing the research and its results and/or public talks given by the team members. However, the impact of this form of outreach cannot easily be analysed and, without large-scale advertising, only reaches a relatively small number of people.  For optical astronomy, numerous student projects have been developed that make use of optical telescopes that are controlled remotely by students  \citep[e.g.][]{mm00,rd07}.  
The majority of these projects have great educational value, but often have little impact on mainstream scientific research. Recently a project was started at the University of Texas, Brownsville (the \textsc{ARCC} project\footnote{http://arcc.phys.utb.edu/ARCC/})  that allows high school students to observe with the Arecibo radio telescope as part of an ongoing pulsar survey (Cordes et al. 2006\nocite{cfl+06}).    A similar project\footnote{http://www.pulsarsearchcollaboratory.com/} allows high school students in West Virginia, U.S.A., to search for pulsars using observations taken with the GreenBank telescope.

The PULSE@Parkes project officially started in December 2007 \citep{hhc+08}.  The project makes use of the Parkes radio telescopes with the aim of 1) obtaining research quality pulsar timing data-sets applicable to many long-term pulsar timing programs, 2) testing remote observing capabilities of the antennae and 3) providing a way to engage many high school students in science.  The students carry out the relatively straightforward observations.  This allows the students to develop their understanding of science, at the same time providing data sets for professional researchers.  In contrast to the ARCC project, the PULSE@Parkes project does not select students based on their current academic standing. PULSE@Parkes also uses pulsar timing observations instead of pulsar survey observations.  A summary of the PULSE@Parkes project can be obtained from our web-site\footnote{http://outreach.atnf.csiro.au/education/pulseatparkes}. 

This paper provides an overview of the project.  A detailed analysis of the educational impact of this project will be published elsewhere. Here, we describe the pulsar sample chosen, describe the new observing techniques, provide initial scientific results and highlight possible results obtainable in the near future.

\section{Observations}

\begin{figure}
\includegraphics[width=7cm,angle=-90]{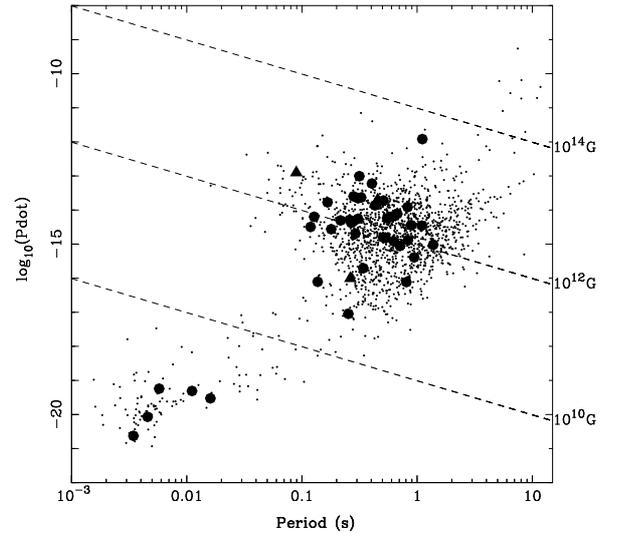}
\caption{A period--period-derivative diagram for the 40 pulsars in the PULSE@Parkes sample observed using the 64-m telescope (filled circles) and the three pulsars observed with the 12-m telescope (filled triangles) overlaid on the known pulsar population.  The dashed lines indicate characteristic magnetic field strengths.}\label{fg:ppdot}
\end{figure}

The PULSE@Parkes observations have been divided into those obtained using the 64-m telescope and those obtained with the 12-m telescope.  The larger telescope provides much higher quality data sets for a large sample of pulsars.  We typically obtain a two-hour observing session per month which yields several observations of a given pulsar over the course of a year.  The 12-m telescope allows far more frequent observations, albeit with poorer sensitivity, allowing no more than a few, bright pulsars to be studied.

\subsection{The 64-m telescope}

\begin{figure*}
\includegraphics[width=14cm]{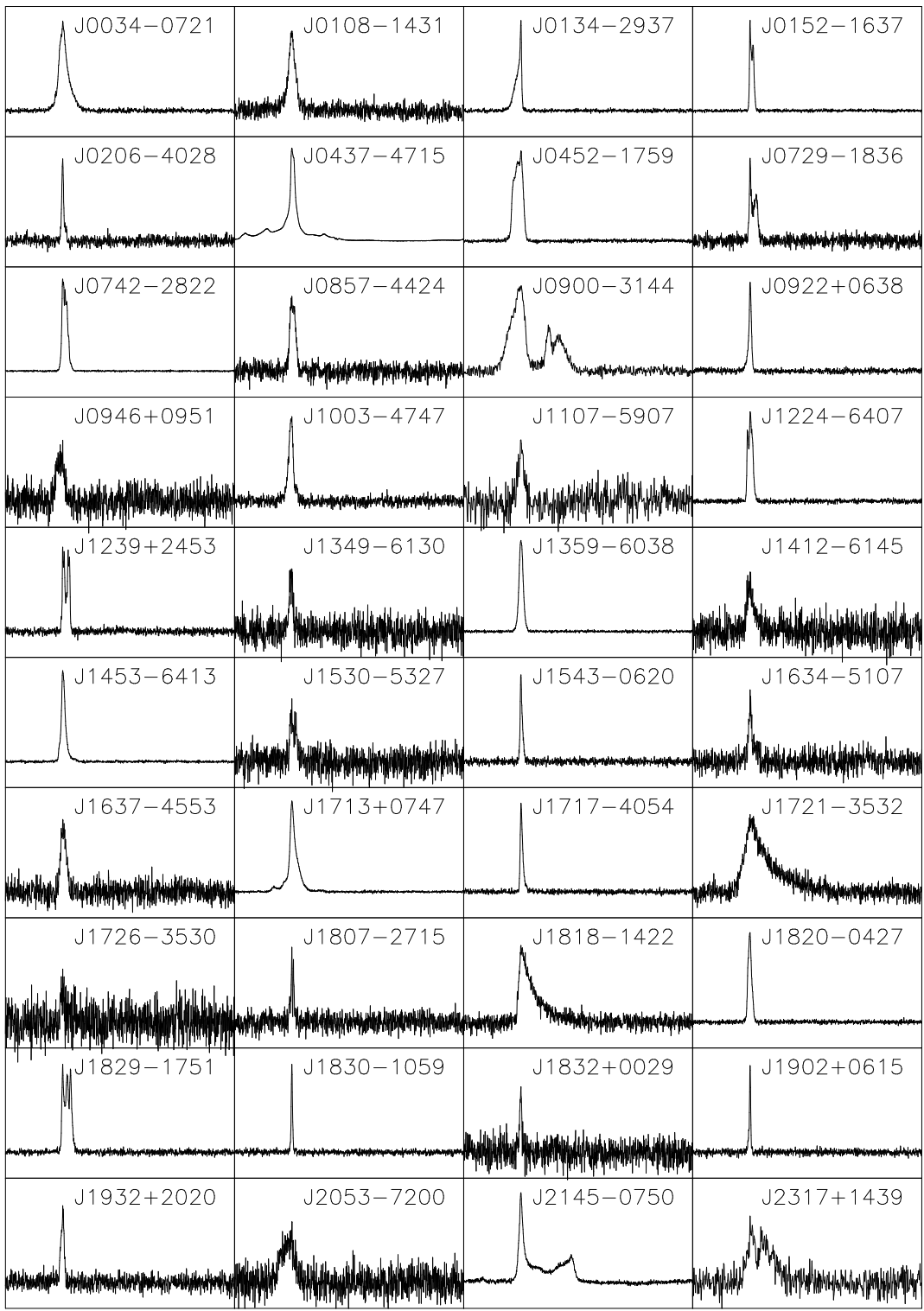}
\caption{Typical pulse profiles (with an arbitrary flux scale) at 20\,cm obtained in a PULSE@Parkes observation.  The highest point in the profile is placed at phase 0.25.}\label{fg:profile}
\end{figure*}

A sample of 40 pulsars has been chosen for observation with the 64-m telescope.  These pulsars were selected 1) because they are useful for long-term timing campaigns, 2)  because they span the entire range of pulsar properties from millisecond pulsars to slow pulsars (their basic spin-properties are summarised in Figure~\ref{fg:ppdot}) and 3) because they provide short, straightforward data sets that can be accessed and analysed by the high school students involved.  In Table~\ref{tb:sample} we list each pulsar's B1950 and J2000 names, a flag indicating whether this pulsar is part of long-term timing programs for gamma-ray (`G') mission support (see \S\ref{sec:gamma}) or for the Parkes Pulsar Timing Array project (`T') (see \S\ref{sec:ppta}). We also indicate whether the pulsar is being monitored to see whether the emission switches `on' and `off' (the pulsar `nulls') in an unexplained manner (`N').  The table also contains the pulsar's pulse-period and its first derivative, its dispersion measure, orbital period and flux density at 1400\,MHz. These parameters were obtained for each pulsar from the Australia Telescope National Facility (ATNF) pulsar catalogue (Manchester et al.  2005\nocite{mhth05}).

The observations are carried out approximately once per month with a new digital filterbank system at an observing frequency close to 1400\,MHz and a bandwidth of 256\,MHz with either the 20\,cm multibeam receiver or the H-OH receiver. Observation lengths are typically 2 to 15\,min depending on the pulsar's flux density.  The data are processed using the \textsc{PSRCHIVE} suite of software (Hotan, van Straten \& Manchester 2004\nocite{hvm04}) that allows polarimetric and flux density calibration.    Pulsar timing models and the resulting timing residuals are obtained using \textsc{tempo2}.

The majority of these observations have been carried out remotely by the high school students from schools listed in Table~\ref{tb:groups} (in this table we list the date of observation, the name of the school, the region containing the school, the approximate number of students that attended, $N_{\rm s}$, and the number of pulsars that were observed, $N_{\rm psr}$).   Currently, the students undertake the observations from the ATNF headquarters in Sydney or from existing science centres (to date, observations have been undertaken from Sydney, the University of Western Australia, the University of Texas at Brownsville and the University of Milwaukee. However, it is expected that we will be able to provide access from science centres around Australia in the near future). This project has become possible by the high-speed network links recently installed between Parkes and Sydney and the installation of web cameras at the observatory that do not create radio interference. The students can access the telescope control software and monitor the observations via web cameras set up outside the telescope and inside the observatory control room.  A professional astronomer at Parkes interacts with the students throughout their observations and is responsible for telescope safety.  The Parkes-based astronomer can immediately take control of the telescope if required.

\begin{table*}
\caption{The PULSE@Parkes sample of pulsars. Uncertainties on parameters are given in parentheses as the error on the last quoted digit.  The pulse-period has been truncated at four decimal places.}\label{tb:sample}
\begin{tabular}{lllllllll}\hline
PSR B      & PSR J 			&Type & $P$ & $\dot{P}$ & DM & $P_b$ & S$_{1400}$ \\
                  &                          		&	   & (s)    & ($10^{-15}$) & (cm$^{-3}$pc) & (d) & (mJy) \\ \hline
B0031$-$07	& J0034$-$0721		& - & 0.9430 & 0.408210(7) & 11.38(8) & - & 11(3)  \\
-		& J0108$-$1431		&G& 0.8075 & 0.070704(12) & 2.38(19) & - & - \\
-		& J0134$-$2937		& - & 0.1369 & 0.0783718(10) & 21.806(6) & -& 2.4  \\
B0149$-$16	& J0152$-$1637		& - & 0.8237 & 1.299202(6) & 11.922(4) & - & 1.5(4)  \\ 
B0203$-$40	& J0206$-$4028		& - & 0.6306 & 1.1994(6) & 12.9(12) & - & 1.0 \\ \\

-		& J0437$-$4715		& T & 0.0057 &0.0000572933(5) & 2.6445(1) & 5.74 & 142(53)\\
B0450$-$18	& J0452$-$1759		& - & 0.5489 & 5.753107(16) & 39.903(3) & - & 5.3(8) \\
B0727$-$18	& J0729$-$1836		& - & 0.5102 & 18.95713(19) & 61.293(10) & - & 1.40(15)  \\
B0740$-$28	& J0742$-$2822		& G &0.1668 & 16.82112(14) & 73.782(2) & - & 15.0(15)  \\ 
-		& J0857$-$4424		& G & 0.3268 & 23.34387(6) & 184.429(4) & - & 0.88(10)  \\ \\

-		& J0900$-$3144		& - & 0.0111 & 0.00004912(13) & 75.702(10) & 18.74 & 3.8(6) \\
B0919$+$06& J0922$+$0638		& - & 0.4306 & 13.71950(3)     & 27.271(6) & - & 4.2(9) \\
B0943$+$10& J0946$+$0951		& - & 1.0977 & 3.49339(20)   & 15.4(5) & - & - \\
B1001$-$47	& J1003$-$4747		& G& 0.3071 & 22.0737(13) & 98.1(12) & - & - \\ 
-		& J1107$-$5907		& N	& 0.2528 & 0.0090(10) & 40.2(11) & - & 0.18  \\ \\

-		& J1224$-$6407		& G& 0.2164 & 4.95332(10)& 97.47(12)& - & 3.9(4) \\
B1237$+$25& J1239$+$2453		& - & 1.3824 & 0.960049(3) & 9.242(6)& - & 10(2) \\	
-		& J1349$-$6130		& G& 0.2593 & 5.125(4) & 284.6(4) & - & 0.58(7) 	\\ 
-		& J1359$-$6038		& NG & 0.1275 & 6.3385(14) & 293.71(14) & - & 7.6(8)\\
-		& J1412$-$6145		& G	& 0.3152 & 98.6598(13) & 514.7(11) & - & 0.47(6) \\ \\

B1449$-$64	& J1453$-$6413		& G & 0.1795 & 2.74610(10) & 71.07(2) & - & 14.0(14)\\
-		& J1530$-$5327		& G& 0.2790 & 4.683(4) & 49.6(10) & - & 0.59(7) 	\\
B1540$-$06	& J1543$-$0620		& - & 0.7091 & 0.87964(4) & 18.403(4) & - & 2.0(7) \\ 
-		& J1634$-$5107		& N& 0.5074 & 1.573(3) & 372.8(20) & - & 0.35 \\
B1634$-$45	& J1637$-$4553		& G & 0.1188 & 3.18790(6) & 193.23(7) & - & 1.10(12) \\ \\
	
-		& J1713$+$0747		& T & 0.0046 & 0.0000085289(5) & 15.9899(6) & 67.83 & 8(2) \\
-		& J1717$-$4054		& N& 0.8877 & - & 308.5(5) & - & 54(5) \\
B1718$-$35	& J1721$-$3532		& G & 0.2804 & 25.18617(8) & 496.0(4) & - & 11.0(11) \\ 
-		& J1726$-$3530		& G & 1.1101 & 1216.751(4) & 727(7) & - & 0.30(4)	\\
B1804$-$27	& J1807$-$2715		& - & 0.8278 & 12.17251(7) & 312.98(3) & - & 0.91(10) \\ \\

B1815$-$14	& J1818$-$1422		& - & 0.2915 & 2.03871(6) & 622.0(4) & - & 7.1(7) \\
B1818$-$04	& J1820$-$0427		& - & 0.5981 & 6.33137(4) & 84.435(17) & - & 6.1(6) \\
B1826$-$17	& J1829$-$1751		& - & 0.3071 & 5.55174(5) & 217.108(9) & - & 7.7(8) \\ 
B1828$-$11	& J1830$-$1059		& G & 0.4050 & 60.02513(9) & 61.50(20) & - & 1.40(15) \\	
-		& J1832$+$0029		& N& 0.5339 & 1.51(20) & 28.3(12) & - & 0.14\\ \\

B1900$+$06& J1902$+$0615		& - & 0.6735 & 7.70638(3) & 502.900(17) & - & 1.10(12) \\
B1929$+$20& J1932$+$2020		& - & 0.2682 & 4.21706(4) & 211.151(11) & - & 1.2(4) \\
B2048$-$72	& J2053$-$7200		& - & 0.3413 & 0.1957(3) & 17.3(4) & - & 6	\\ 
-		& J2145$-$0750		& T& 0.0161 & 0.000029757(3) & 9.0031(2) & 6.84 & 8(2)\\
-		& J2317$+$1439 		& - & 0.0034 & 0.00000242(3) & 21.907(3) & 2.45 & 4(1) \\
\hline
\end{tabular}
\end{table*}

\begin{table*}
\begin{center}\caption{Student groups that have taken part in the PULSE@Parkes project}\label{tb:groups}
\begin{tabular}{lp{7cm}lrr}\hline
Date & School & Region$^\dagger$ & $N_{\rm s}$ & $N_{\rm psr}$ \\ \hline
 04/12/07 & Kingswood High School & Syd. & 10 & 6 \\
 13/02/08 & Muswellbrook High School & N.S.W. & 8 & 10 \\
 19/03/08 & Duncraig Senior High School, Mt Lawley Senior High School and Willetton Senior High School & W.A. & 15 & 12\\
 22/05/08 & Mt Carmel High School & Syd. & 26 & 9 \\
 27/06/08 & Caroline Chisholm College & Syd. & 9 & 8 \\
22/07/08  & Asquith Girls' High School & Syd. & 14 & 4\\
13/08/08  & Ambarvale High School & Syd. & 11 & 9 \\
01/09/08  & Epping Boys' High School & Syd. & 11 & 8\\
23/09/08 & Lithgow High School & N.S.W. & 12 & 9 \\ 
28/10/08 & Casimir Catholic College & Syd. & 10 & 7 \\
28/10/08 & Brownsville, Texas & U.S.A. & 8 & 6 \\
13/11/08 & International Grammar School & Syd. & 10 & 9 \\
02/12/08 & Terrigal High School & Syd. & 16 & 8 \\
02/12/08 & Brownsville, Texas & U.S.A. & 8 & 10 \\
11/02/09 & Hurlstone Agricultural High School & Syd. & 16 & 7 \\
17/03/09 & New England DET Aboriginal G\&T & N.S.W. & 3 & 8 \\
12/04/09 & Penrith Anglican College & Syd. & 17 & 8 \\
\hline\end{tabular}

$^\dagger$ the regions included are the United States of America (U.S.A.), the Sydney region (Syd.), New South Wales not counting the Sydney region (N.S.W.) and Western Australia (W.A.)
\end{center}
\end{table*}

\begin{figure}
\includegraphics[width=7cm,angle=-90]{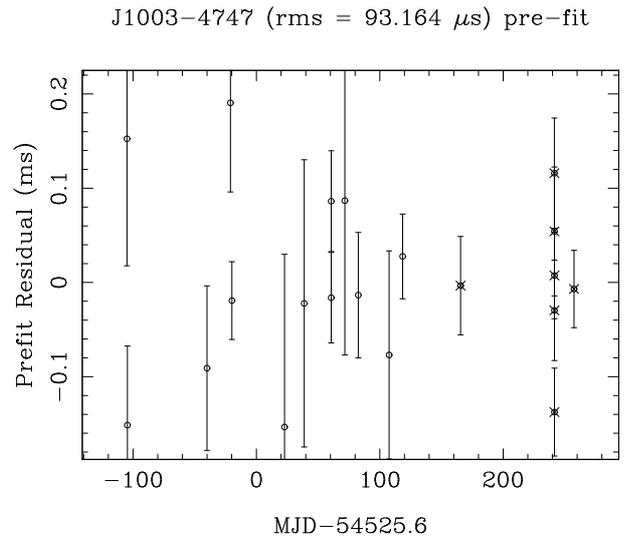}
\caption{Timing residuals for PSR~J1003$-$4747. The left-most points were obtained with the second generation digital filterbank system (PDFB2) and the more recent observations (with cross symbols) with the third generation system (PDFB3).}\label{fg:1003}
\end{figure}

In Figure~\ref{fg:profile} we plot a typical total intensity folded pulse profile for each of the 40 pulsars at an observing frequency close to 1.4\,GHz. The signal-to-noise ratio of the profiles is typical of that obtained in a PULSE@Parkes observing session.   For some pulsars we now have more than a year's worth of monthly observations. For other pulsars we currently have significantly reduced sampling.  An example of the timing achieved is shown in Figure~\ref{fg:1003} for PSR~J1003$-$4747.  

\subsection{The 12-m diameter telescope}

\begin{figure*}
\includegraphics[width=4cm,angle=-90]{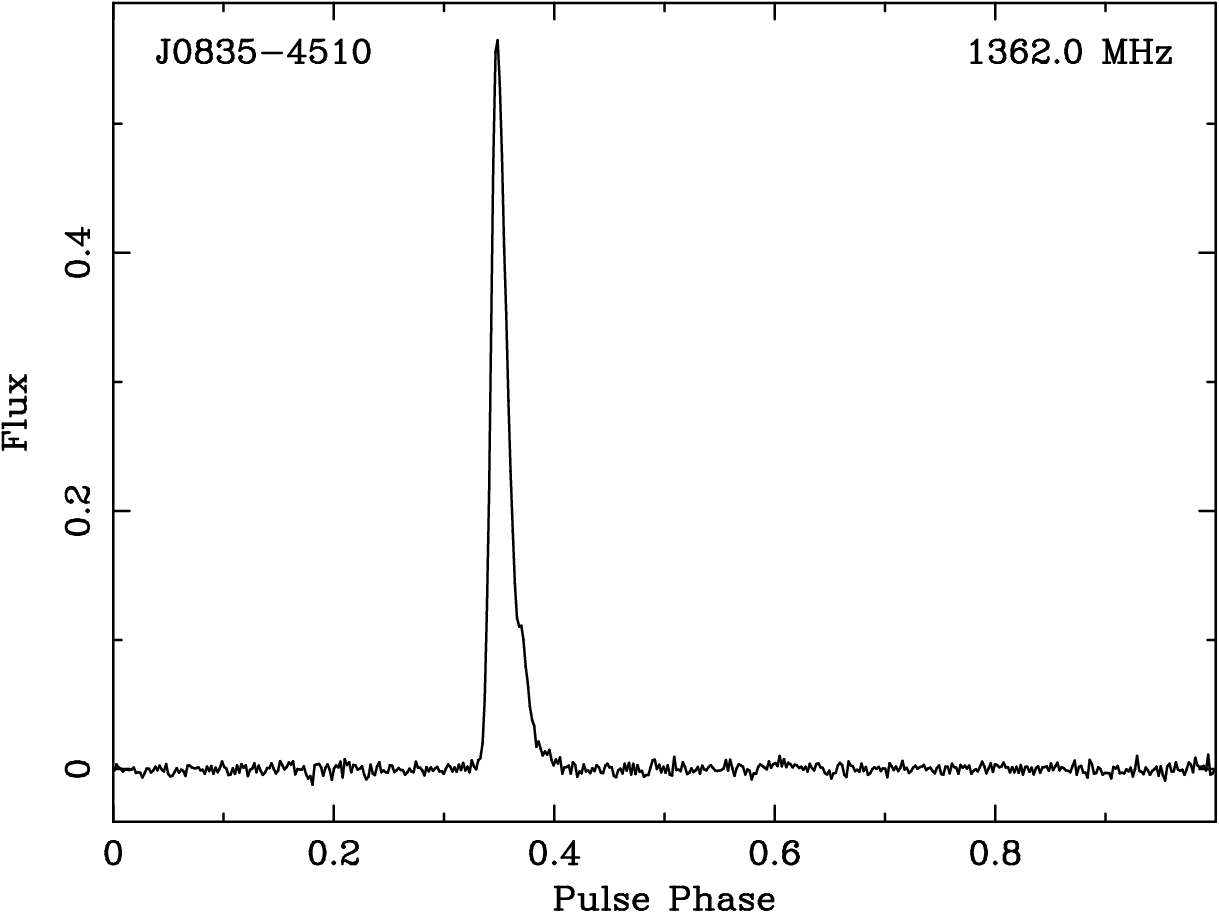}
\includegraphics[width=4cm,angle=-90]{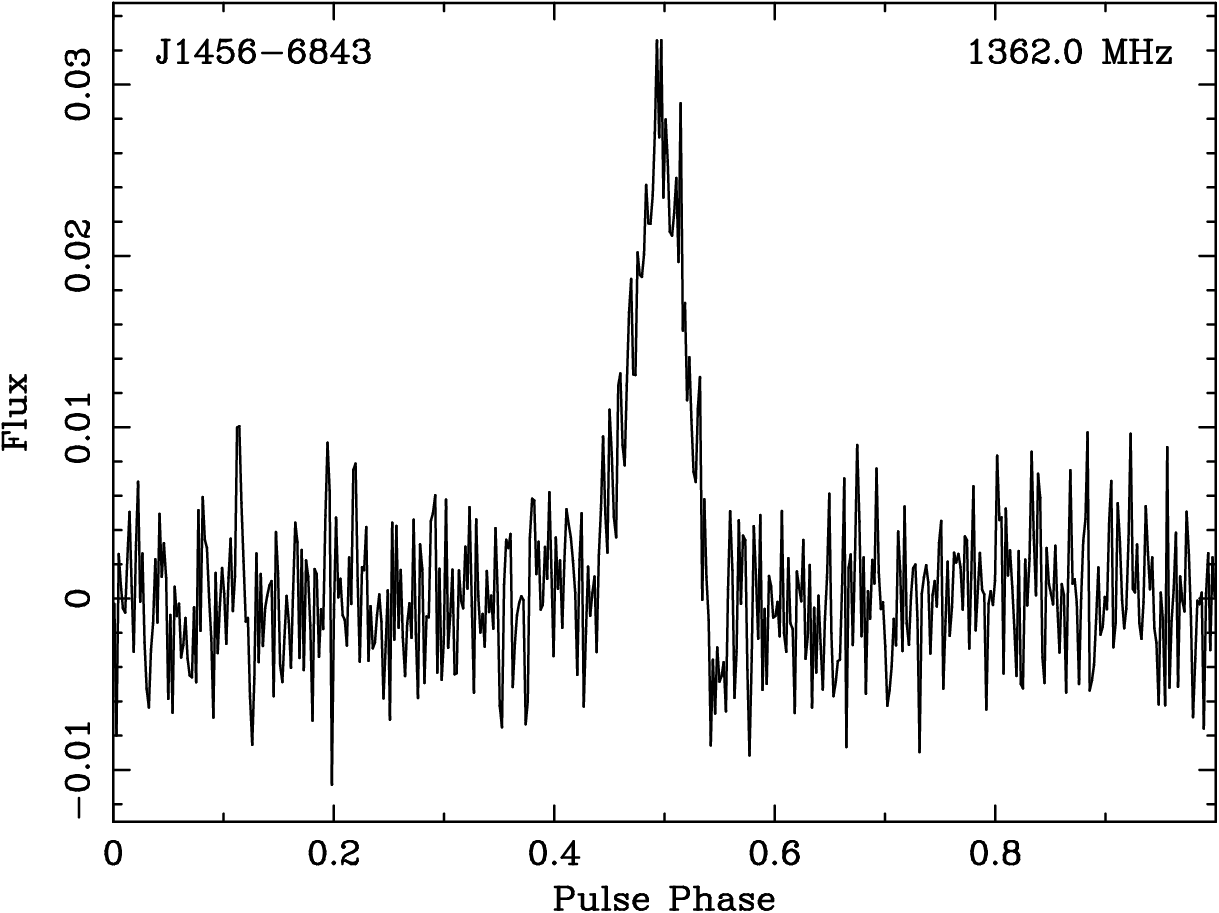}
\includegraphics[width=4cm,angle=-90]{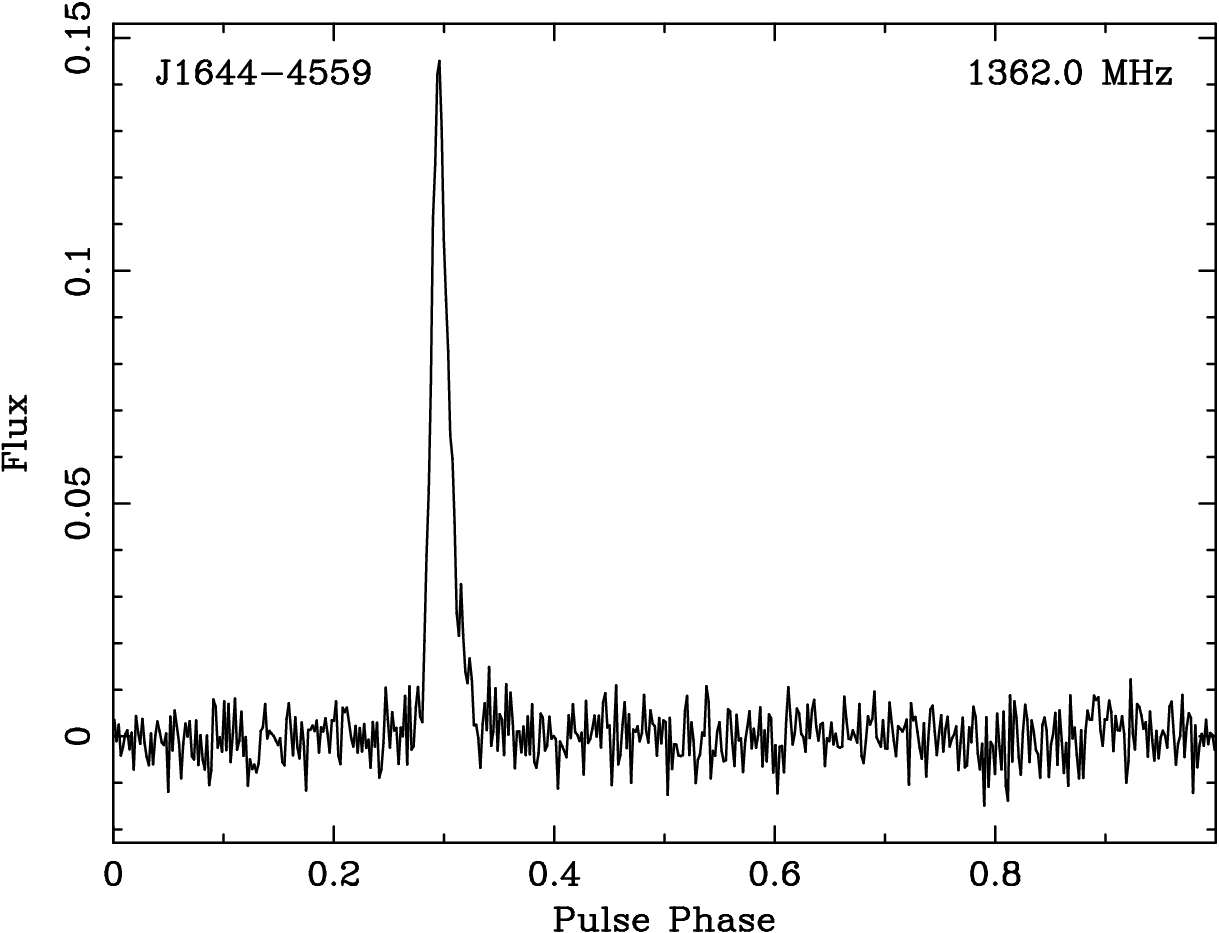}
\caption{Pulse profiles for PSRs~J0835$-$4510, J1456$-$6843 and J1644$-$4559 obtained using the Parkes 12-m antenna.}\label{fg:12m_profiles}
\end{figure*}

The 12-m telescope at Parkes is a prototype antenna for the Australian Square Kilometre Array Pathfinder telescope (ASKAP) and is being used for engineering testing of phased array feeds.  The PULSE@Parkes project has obtained the first scientific data using this telescope which has been used to test and characterise the antenna. All the data described in this paper that were taken with this telescope were obtained using a simple feed providing a 20\,MHz bandwidth around an observing frequency of 1.4GHz, with a system temperature $\sim 150$K.  As a backend system we use a pulsar digital filterbank that records two polarisations with 1024 frequency channels.  We store a folded pulse profile every 30\,seconds and observe a typical pulsar for many hours.   We present the basic properties of the three pulsars currently observed by this system in Table~\ref{tb:12m}.   The telescope is controlled remotely using high-speed network links in a manner similar to the 64-m telescope although the observations are completely remote and do not require an astronomer at the Parkes site.    In Figure~\ref{fg:12m_profiles} we show the typical pulse profiles achieved in a single observation using the 12-m telescope.

\begin{table*}
\begin{center}\caption{The PULSE@Parkes sample of pulsars observed using the Parkes 12-m antenna.}\label{tb:12m}
\begin{tabular}{llllll}\hline
PSR B      & PSR J 			& $P$ & $\dot{P}$ & DM & S$_{1400}$ \\ 
                  &                		         & (s)    & ($10^{-15}$) & (cm$^{-3}$pc) & (mJy) \\ \hline
 B0833$-$45 & J0835$-$4510 & 0.0893 & 125.008(16) & 67.99(1) & 1100.0 \\
 B1451$-$68 & J1456$-$6843 & 0.2634 & 0.09826(13) & 8.6(2) & 80.0 \\
 B1641$-$45 & J1644$-$4559 & 0.4551 & 20.0902(6) & 478.8(8) & 310.0 \\
\hline\end{tabular}\end{center}
\end{table*}

\section{Scientific aims and initial results}

The main scientific aims of the PULSE@Parkes project are to study 1) the long-term timing stability of radio pulsars, 2) glitch events, 3) pulse profile stability over long time scales and 4) the extreme nulling phenomenon.  To achieve these aims data spanning many years are required. Despite only running the program for one year we have already achieved some results which we present in this section.

\subsection{PSR~J0835$-$4510}

\begin{figure}
\includegraphics[width=7cm,angle=-90]{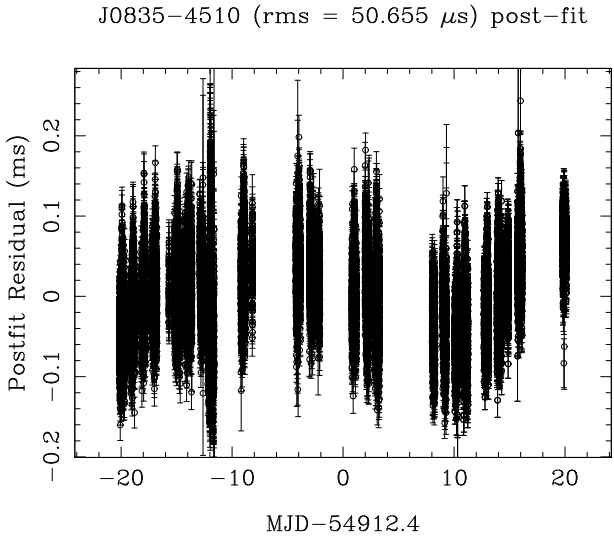}
\caption{Timing residuals for the Vela pulsar.}\label{fg:vela}
\end{figure}

The Vela pulsar, PSR~J0835$-$4510, has been regularly observed since its discovery \citep{lvm68}.  Between the years 1981 and 2005 this pulsar was monitored using the Mount Pleasant observatory \citep{dlm07}.   Approximately 16 glitches have been reported that occurred over 20\,yrs of observing. We are now continuing to monitor this pulsar using the Parkes 12-m antenna and expect to detect one glitch approximately every 2\,yrs. Our sampling should allow us to determine the time of the glitch (if the glitch occurs while the pulsar is observable from the Parkes Observatory) to within $\sim 30$\,s and to study the post-glitch recovery in detail.   In Figure~\ref{fg:vela} we show the timing residuals obtained over 40\,d of observing.  No glitch has so far been discovered in our data.

\subsection{PSR~J1107$-$5907}

PSR~J1107$-$5907 was discovered in the Parkes Multibeam Pulsar Survey \citep{lfl+06} and has been shown to exhibit three states of emission \citep{okl+06}.  In one of these states the pulsar is completely `off' and no pulsed radio emission is detectable. In the second state the pulse profile is weak and has a narrow profile.  In the third state the pulsed signal is extremely strong and has a wide profile. The time scales and properties of these different states still need to be determined.  We currently have ten observations of this source.  For these observations it was undetectable seven times and in the weak state for the remaining three.  We have not detected this pulsar in its strong state.

\subsection{PSR~J1717$-$4054}

\begin{figure*}
\includegraphics[width=6cm,angle=-90]{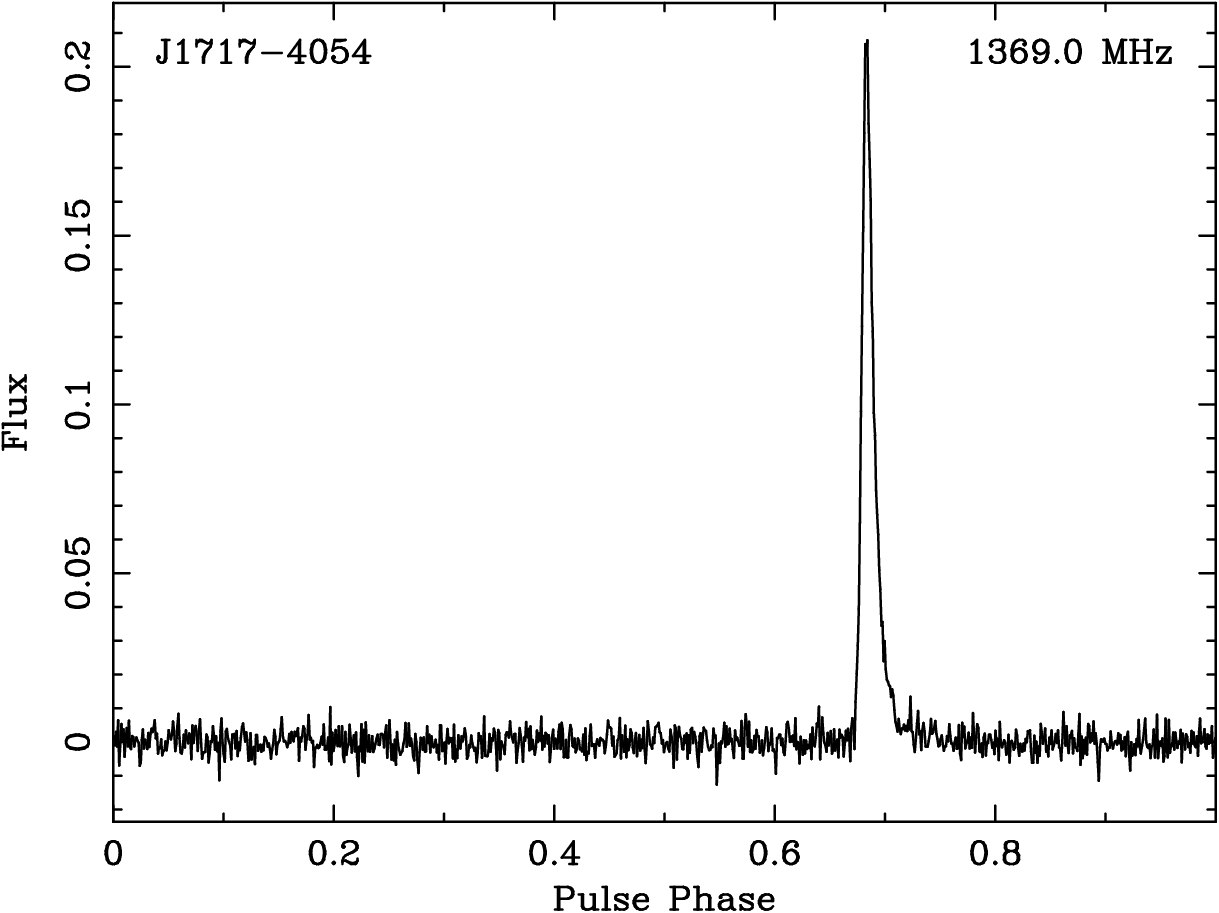}
\includegraphics[width=6cm,angle=-90]{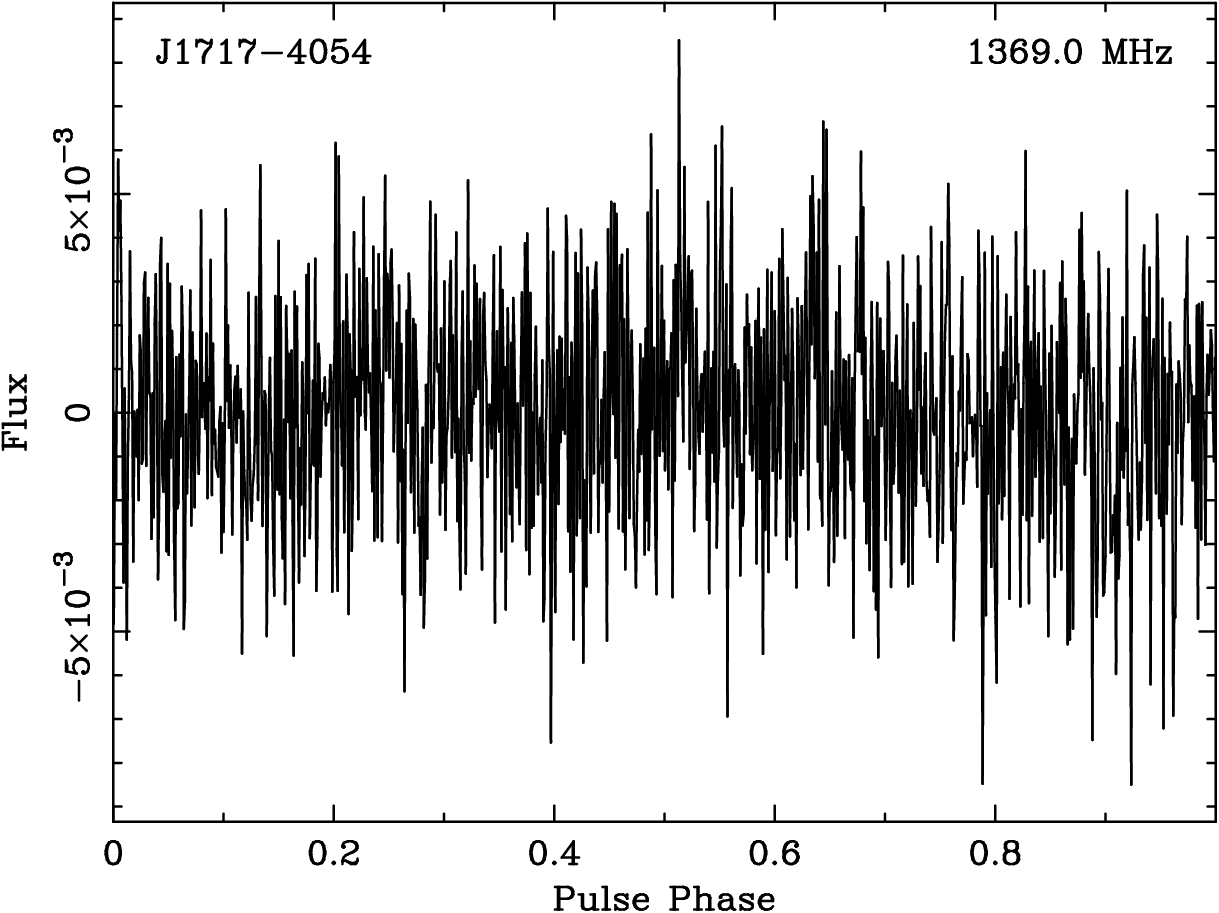}
\caption{Two observations of PSR~J1717$-$4054.  The right-hand panel contains an observation taken 15 minutes after the completion of the observation shown in the left-hand panel.}\label{fg:1717}
\end{figure*}

PSR~J1717$-$4054 was discovered in the Parkes Southern Survey \citep{jlm+92} and redetected in the Pulsar Multibeam Pulsar Survey. This source has been described by O'Brien et al. (2006)\nocite{okl+06} and is found to `switch-off' for long periods of time.  O'Brien et al. (2006) reported that it was `on' for only $\sim 20$\% of the time.  However, this source has not yet been studied in detail.  We have attempted to observe this source four times and seen it twice. In Figure~\ref{fg:1717} we show two observations of the source taken 15 minutes apart by students from Lithgow High School in Australia.  In the first observation the pulsar is `on' and is bright within one minute of observing.  However, in the second observation the pulsar has switched `off' and is completely undetectable even after observing for five minutes.  

\subsection{Support for gamma-ray missions}\label{sec:gamma}

14 of the PULSE@Parkes pulsars are also being observed as part of the Parkes gamma-ray support project.  The PULSE@Parkes observations are automatically being processed with the gamma-ray data processing pipeline and the resulting observations and timing solutions being supplied to the Fermi and AGILE mission teams for folding of the gamma-ray photons.  Many of these pulsars show large period variability implying that it is difficult to obtain a coherent timing solution over many weeks or months.  The PULSE@Parkes observations improve the data sampling for these pulsars.

\subsection{Support for the Parkes Pulsar Timing Array project}\label{sec:ppta}

The Parkes Pulsar Timing Array (PPTA) project began in the year 2003.  The three main aims of this project are 1) to detect gravitational wave signals, 2) to establish a pulsar-based time scale for comparison with terrestrial time standards and 3) to improve the Solar system planetary ephemeris. An overview of the project has been published by Manchester (2008) and recent details of the gravitational wave aspects of the project by Hobbs et al. (2009).  This project requires numerous observations of very stable millisecond pulsars in order to achieve the aims.  The PULSE@Parkes project contains three pulsars (PSRs~J0437$-$4715, J1713$+$0747 and J2145$-$0750) that are part of the PPTA project and one further pulsar, PSR~J2317$+$1439 that is being studied for possible future inclusion in the PPTA sample. The observations of these pulsars made by the high school students are automatically  processed as part of the PPTA pipeline and used in the resulting data sets. The observations are also used in order to test the hardware, software and calibration methods used for the PPTA project.

\section{Educational value}

PULSE@Parkes aims to engage high school students in science by providing them with a stimulating example of real science using a major national facility. The students interact directly and via video conferencing with active research scientists. Within the context of radio astronomy the students develop skills in science, information and communication technology, problem solving and group work.  Prior to their observing session, a school visit takes place in which the students and their teachers are given an overview of radio astronomy, pulsars and basic astronomy.   At the start of the observing sessions the students are given a short talk which includes detailed information on how they use the telescope control software and a short movie that provides a virtual tour of the telescope.  During their observations the students are exposed to real observational issues that affect the quality of data and results.  They are expected to keep an online log that describes the pulsars being studied, impulsive radio interference and any hardware failures.  Each group of students observes a few pulsars. The data they collect are immediately available online allowing the students to carry out small projects.  Under supervision from a professional astronomer the students determine the distance to their pulsars from the pulsars' dispersion measures. New projects are being developed that will allow the students to compare their observations with previous data in order to estimate pulsar ages and search for glitch events. Post-observing collaborations among students from different schools will be encouraged in these new projects. There is
also opportunity for the students to explore the topic further in their school science classs. For example, under the guidance of their teacher, they could develop their own investigations for use in programs such as the various state science investigation awards.  Currently the program is offered to students in Years 10 - 12, is  open to any school to apply and is non-academically selective. A  typical two-hour observing session involves up to 24 students, split into groups of four.

PULSE@Parkes is the first stage in developing, testing and implementing educational projects planned for larger-scale schemes in the future using new facilities such as ASKAP and the Square Kilometre Array (SKA). Given that there has been much less work on high school student use and understanding of radio astronomy compared with optical astronomy, a key component of the project has been to conduct educational research to evaluate the scheme and guide ongoing development.   The results of this research will shortly be published in appropriate science education research journals.

\section{Conclusion}

PULSE@Parkes is a new, international project with both scientific and educational impact. It demonstrates a new observing model that shares with high school students the excitement of carrying out scientific observations using state-of-the-art instrumentation, while providing essential observations for cutting-edge science.

The project uses the high-speed networking systems now available to allow remote observations to be carried out at the Parkes telescope and demonstrates the manner in which educational programs and scientific studies can be linked together.   The PULSE@Parkes model is already being considered for carrying out other experiments, such as pulsar or neutral hydrogen surveys, in the future.

\section*{Acknowledgments} 

We acknowledge the many scientific, education and engineering staff in Sydney and Parkes needed to make the PULSE@Parkes project a reality. The Parkes radio telescope is part of the Australia Telescope which is funded by the Commonwealth of Australia for operation as a National Facility managed by CSIRO.  GH is funded by a QEII fellowship awarded by the Australian Research Council. We thank the Fermi and Parkes Pulsar Timing Array teams for being supportive of the PULSE@Parkes project.
We thank M. Kesteven for his support in using the 12-m antenna at Parkes.  We gratefully acknowledge the students and teachers of the schools listed in Table~\ref{tb:groups} that have taken part in the project.

\bibliography{modrefs,psrrefs,crossrefs,journals}
\bibliographystyle{mn2e}



\end{document}